\documentclass[11pt, letter]{article}
\usepackage{graphicx} % Required for inserting images

\usepackage{algorithmic}
\usepackage{algorithm}

\usepackage{amsthm}
\usepackage{amsmath}
\usepackage{amssymb}

\newtheorem{theorem}{Theorem}[section]
\newtheorem{lemma}[theorem]{Lemma}
\theoremstyle{definition}

\title{Quantum Algorithm with Quadratic Speedup for 2D Pattern Matching}
\author{Rishabh Gala\\Baton Rouge Magnet High School\\Baton Rouge, LA 70806, USA\\email:rishabhrgala@gmail.com
\and
Rahul Shah\\Louisiana State University\\Baton Rouge, LA 70803, USA\\email:rahul@lsu.edu}
\date{}

\begin{document}

\maketitle

\begin{abstract}

We consider the problem of two-dimensional pattern matching under the quantum query model. Here, we are given a rectangular text $T[0 \dots n_1-1][0 \dots n_2-1]$ and a pattern $P[0 \dots m_1-1][0 \dots m_2-1]$. Let $N = n_1 n_2$ and $M = m_1 m_2$ denote the sizes of the text and the pattern, respectively, where $M < N$. Building upon the quantum advantage provided by Grover's search (and subsequent D\"urr--H{\o}yer minimum finding), we present an algorithm that requires $\widetilde{O}(\sqrt{N})$ query complexity to determine whether $P$ occurs in $T$.

Our solution employs deterministic sampling as a primary component. As noted in earlier literature, deterministic sampling alone is inadequate for 2D matching, making witness tables essential. In our setting, we assume that the witness table of the pattern is precomputed classically and provided as an oracle to the quantum algorithm. Our main novel contribution is demonstrating how to compute and exploit a geometric structure known as the \emph{skyline} without forfeiting quantum advantage.
\end{abstract} 

\section{Introduction}

Two-dimensional pattern matching is one of the most fundamental and intriguing problem areas in string processing~\cite{amir1994two}. The multidimensional nature of periodicity poses unique structural challenges for exact matching, index construction, and space-efficient text compression across various computing models. In this work, we study 2D pattern matching within the quantum query model. We show that one can achieve a near-quadratic quantum speedup for searching a rectangular pattern $P[0 \dots m_1-1][0 \dots m_2-1]$ inside a rectangular text $T[0 \dots n_1-1][0 \dots n_2-1]$, where $m_1 \le n_1$, $m_2 \le n_2$, $M = m_1 m_2$, and $N = n_1 n_2$.

The substantial increase in attention and funding for quantum computing over the last decade has sparked a renewed push toward developing efficient quantum algorithms. Within string matching, a diverse array of quantum results has recently emerged~\cite{wangying2020, jinnogler2024, gibneyjin2024, gibneythankachan2023, montanaro2017, akmaljin2023, legallseddighin2022}. Much like classical models of computation—such as PRAM, External Memory, and Cache-Oblivious models—the quantum query model introduces unique structural constraints. These constraints restrict the direct application of traditional algorithmic tools without forfeiting quantum advantage, necessitating the invention of novel techniques tailored specifically to quantum computational paradigms.

At its core, our approach builds upon the foundational framework for one-dimensional quantum pattern matching introduced by Ramesh and Vinay~\cite{ramesh2003string} and Wang and Ying~\cite{wangying2020}, which in turn adapted the classical deterministic sampling paradigm proposed by Vishkin~\cite{vishkin1991optimal} for fast parallel (PRAM) algorithms. However, relying solely on deterministic sampling presents significant structural hurdles in 2D matching, prompting subsequent classical algorithms to leverage witness tables (see~\cite{amir1992, amir1994, galilpark1992, crochemore1993ipl, crochemore1998, amirbenson1992periodicity, amirbenson1998} for this line of work). Due to these structural barriers, our quantum algorithm assumes access to a precomputed witness table that explicitly identifies character positions where two shifted pattern candidates differ.

The fundamental distinction between 1D and 2D pattern matching lies in shift geometry: while 1D overlaps are strictly collinear, 2D candidate shifts can occur along arbitrary 2D vectors. Consequently, 2D periodicity branches into 1D-linearly periodic and 2D-lattice periodic regimes. A naive approach would reduce 2D matching to multiple 1D instances by analyzing the deterministic sample and periodicity of individual rows; however, such a reduction yields a suboptimal quantum query complexity of $\widetilde{O}(N^{3/4})$. Our target is to achieve the near-optimal subquadratic query bound of $\widetilde{O}(\sqrt{N})$.

Central to our solution is the \emph{skyline} structure and the creative deployment of Grover's and D\"urr--H{\o}yer search~\cite{durr1996quantum, grover1996fast} to construct this skyline in 2D geometry. We introduce a recursive, area-based divide-and-conquer partitioning scheme to build the skyline efficiently within the quantum model. Although classical precomputation of the witness table requires $O(M)$ time—resulting in a hybrid quantum (or qRAM-based) algorithm with an overall operational time complexity of $O(M + \sqrt{N})$—we note that in many realistic problem settings, $M \le O(\sqrt{N})$.

Our pipeline proceeds as follows: we first use deterministic sampling to locate candidate sample matches within text blocks of size $(m_1 + m_1/(2k)) \times (m_2 + m_2/(2k))$ for a fixed constant $k$. Depending on the periodicity regime, we reduce the number of candidate matches by a factor of $\sqrt{M}$, enabling us to afford sequential witness duels to prune the candidate set further. Finally, we deploy skyline-based verification to determine whether the pattern occurs within the block. Combining this block verification routine with an outer quantum search across all blocks yields an overall complexity of $\widetilde{O}(\sqrt{N})$.

\smallskip
\noindent\textbf{Map.}
The remainder of this paper is organized as follows. Section~2 defines the problem and establishes preliminaries. Sections~3 and~4 detail the deterministic sampling scheme, discuss the structural intricacies introduced by 2D periodicity, and demonstrate how witness duels reduce the candidate set to the desired structure and density. Section~5 introduces the skyline framework and presents its $\widetilde{O}(\sqrt{N})$ construction algorithm. Section~6 describes the final verification phase that pinpoints actual pattern occurrences from the pruned candidate set. Finally, Section~7 concludes with a formal quantum query and time complexity analysis.

\section{Preliminaries}
\label{sec:prel}
For pattern matching, we consider a rectangular pattern $P[0 \ldots m_1-1][0 \ldots m_2-1]$ embedded in a rectangular text $T[0 \ldots n_1-1][ 0 \ldots n_2-1]$. Let $M = m_1 \cdot m_2$ denote the size of the pattern (i.e.\ its number of characters) and $N = n_1 \cdot n_2$ the size of the text. Let $k$ be a constant to be fixed later.

A \emph{shift} $(a,b)$, denoted $P_{(a,b)}$, of the pattern $P[0 \ldots m_1-1][0 \ldots m_2-1]$ is the placement of $P$ with its top-left corner at $(a,b)$:
\[
P_{(a,b)}[i,j] \;=\; P[i-a][j-b], \qquad \text{with index space } P_{(a,b)}[a \ldots a+m_1-1][ b \ldots b+m_2-1].
\]

A \emph{column} at position $(i,j)$, with $i < m_1 + m_1/k$ and $j < m_2 + m_2/k$, is the list of all shifts that are ``pierced'' by that position. Formally,
\[
S_{i,j} \;=\; \bigl\{\, ((a,b);\, c) \;:\; P[i-a][ j-b] = c \;\land\; a \le i < a+m_1 \;\land\; b \le j < b+m_2 \,\bigr\}.
\]

If $m_1/k \le i < m_1$ and $m_2/k \le j < m_2$, then column $(i,j)$ pierces every shift under consideration. Such a column is called a \emph{middle column}.

The \emph{middle region} of a pattern $P[0 \ldots m_1][0 \ldots m_2]$ is
\[
P^{\mathrm{mid}} \;=\; P[\,m_1/k \ldots m_1 - m_1/k][ m_2/k \ldots m_2 - m_2/k\,].
\]
%is defined as the subset of a pattern with indices $(x,y)$ satisfying $m_1/k \le x \le m_1 - m_1/k$ and $m_2/k \le y \le m_2 - m_2/k$.
The \emph{inner middle region} is defined analogously, with the factor of $m_i/k$ quadrupled on each side:
\[
P^{\mathrm{inmid}} \;=\; P[\,4m_1/k \ldots m_1 - 4m_1/k][ 4m_2/k \ldots m_2 - 4m_2/k\,].
\]

$k$ is held constant at a value at least $10$. This ensures the dimensions of the inner middle region are at least $2m_1/k \times 2m_2/k$.
%of a pattern with indices $(x,y)$ satisfying $2m_1/k \le x \le m_1 - 2m_1/k$ and $2m_2/k \le y \le m_2 - 2m_2/k$.

\section{Deterministic Sampling}
\subsection{Deterministic Sampling for an Aperiodic Pattern}

We first construct a deterministic sample across the shifts $(a,b)$ with $0 \le a \le \lfloor m_1/k \rfloor$ and $0 \le b \le \lfloor m_2/k \rfloor$.

Deterministic sampling proceeds iteratively. At every point during the procedure we maintain a set $\mathcal{S}$ of \emph{surviving} shifts, initialized as
\[
\mathcal{S} \;=\; \bigl\{\, (a,b) \;:\; 0 \le a \le \lfloor m_1/k \rfloor \;\land\; 0 \le b \le \lfloor m_2/k \rfloor \,\bigr\}.
\]
As columns are appended to the deterministic sample, any shift that fails to match every column currently in the sample is removed from $\mathcal{S}$.

At each iteration, use Dürr–Høyer search to pick any two distinct surviving shifts $s_1, s_2 \in \mathcal{S}$. (Any procedure that returns two distinct surviving shifts suffices; for instance, one can impose any total order on $\mathcal{S}$ and take the Dürr–Høyer minimum and maximum.) If no two distinct shifts remain --- i.e.\ $|\mathcal{S}| \le 1$ --- deterministic sampling is complete.

Otherwise, use Grover's search to find a middle column at which $s_1$ and $s_2$ disagree. If no middle column is found, then since some column must exist where $s_1$ and $s_2$ disagree due to the pattern being aperiodic, the period breaker algorithm can be utilized. Append this column to the deterministic sample, labeled with either of the two conflicting characters, chosen uniformly at random. Each such step removes at least one of $s_1$ or $s_2$ from $\mathcal{S}$, and with probability at least $1/2$ removes at least half of the remaining shifts, so the expected iteration count and deterministic sample size is logarithmic.

\subsection{Deterministic Sampling for a Pattern with One Period (1-D Periodic)}

Let the pattern's period be $v = (x_v, y_v)$.

As before, we construct a deterministic sample across the shifts $(a,b)$ with $0 \le a \le \lfloor m_1/k \rfloor$ and $0 \le b \le \lfloor m_2/k \rfloor$, maintaining a set $\mathcal{S}$ of surviving shifts initialized as
\[
\mathcal{S} \;=\; \bigl\{\, (a,b) \;:\; 0 \le a \le \lfloor m_1/k \rfloor \;\land\; 0 \le b \le \lfloor m_2/k \rfloor \,\bigr\}.
\]
Again, any shift that fails to match a column added to the deterministic sample is removed from $\mathcal{S}$.

At each iteration, use Dürr–Høyer search to find a surviving shift $s_l = (x_l, y_l)$ (for instance the one with lowest $x$-coordinate, tiebroken by highest $y$). Then use Grover's search to find another surviving shift $s_o = (x_o, y_o)$ such that $s_o - s_l$ is \emph{not} an integer multiple of the period $v$; that is, there is no integer $n$ with $s_o - s_l = n\,v$.

If no such $s_o$ exists, deterministic sampling is complete. Otherwise, use Grover's search to find a middle column at which $s_l$ and $s_o$ disagree; if none exist, since there must exist some column where they differ as $v$ is the only period of the pattern, the period breaker algorithm can be utilized. Append this column with one of the two characters from $s_l$ or $s_o$ chosen uniformly at random. As in the aperiodic case, each such step removes with probability at least $1/2$ at least half of the remaining shifts, so the expected iteration count and deterministic sample size is logarithmic.

\subsection{Deterministic Sampling for a Pattern With Two (or more) Distinct Periods}

In this case, we can assume $v_1$ and $v_2$ to be 2 distinct periods of the pattern, found from the witness table.

\begin{lemma}
The middle region of the pattern is lattice periodic with respect to $v_1$ and $v_2$.
\end{lemma}
\begin{proof}
(proof). Take two arbitrary indices $(a,b)$ and $(c,d)$ in the middle region, and let $(c-a,\,d-b)=\alpha v_1+\beta v_2$.
Note that the line connecting $(a,b)$ and $(c,d)$ must lie completely within the middle region.
Create a list of indices of the form $(a,b)+c_1v_1+c_2v_2$, starting with $(a,b)+0v_1+0v_2$ and ending with $(a,b)+\alpha v_1+\beta v_2=(c,d)$.
For each element $(a,b)+c_1v_1+c_2v_2$, if $\frac{c_1}{c_2}<\frac{\alpha}{\beta}$, then make the next element $(a,b)+(c_1+1)v_1+c_2v_2$;
otherwise, make it $(a,b)+c_1v_1+(c_2+1)v_2$.
Every element must be at most $v_1$ or $v_2$ away from the central line, and therefore lies in the pattern.
Since every element of the list is separated from the last by a period, all store the same character, so $P[a][b]=P[c][d]$.
\end{proof}

\begin{lemma}
If there exists distinct linear combinations of $v_1$ and $v_2$, $v_3$ and $v_4$, such that either one has an $x$ or $y$ component of 0, or they lie in opposite quadrants, and both are bound by dimensions $\frac{3m_1}{k} \times \frac{3m_2}{k}$, then the entire pattern is lattice periodic with respect to $v_1$ and $v_2$.
\end{lemma}

\begin{proof}
We assume $k\ge 8$ throughout.
We must show that for any $(a,b)$ and $(c,d)$ separated by $\alpha v_1$ and $\beta v_2$, that $P[a][b]$ must equal $P[c][d]$.
Note that since the middle region of the pattern is proven to be lattice periodic, it is sufficient to prove
$P[a][b]=P[a'][b']$, where
$a'=a+\lambda_1x_{v_1}+\lambda_2x_{v_2}$ and $b'=b+\lambda_1y_{v_1}+\lambda_2y_{v_2}$
for some integers $\lambda_1,\lambda_2$ such that $(a',b')$ lies within the middle region.
This is because, by the same proof, $P[c][d]$ can also be linked to an index in the middle region, so then by lattice periodicity, $P[a][b]$ must equal $P[c][d]$.
 
Since $v_3$ and $v_4$ are linear combinations of $v_1$ and $v_2$, it is sufficient to show that after some applications of $v_3$ and $v_4$
(or their negative versions), $(a,b)$ will be transformed to a point within the middle region.
This is because after every application of these periods, the new index must store the same character in $P$ as the previous.
 
\medskip
\noindent\textbf{Case 1: $v_3$ or $v_4$ is horizontal or vertical.}
\medskip
 
Without loss of generality, let $v_3$ be horizontal, and since $v_4$ is distinct, it must have a non-zero $y$ component.
Without loss of generality, let $v_4$ be in quadrant 4 or be vertical with negative $y$ component.
 
To map $(a,b)$ to a cell in the middle region, attempt to first ensure its $y$ coordinate is within $\bigl[\frac{m_2}{k},\,m_2-\frac{m_2}{k}\bigr]$.
If $b<\frac{m_2}{k}$, then repeatedly do the following: repeatedly add $v_3$ until $a>\frac{m_1}{2}$, then subtract $v_4$.
If $b>m_2-\frac{m_2}{k}$, then repeatedly do the following: repeatedly subtract $v_3$ until $a<\frac{m_1}{2}$, then add $v_4$.
Once $\frac{m_2}{k}<b<m_2-\frac{m_2}{k}$, $a$ can be put within $\frac{m_1}{k}$ and $m_1-\frac{m_1}{k}$ by repeatedly adding or subtracting $v_3$.
 
\medskip
\noindent\textbf{Case 2: $v_3$ and $v_4$ lie in orthogonal quadrants, with neither being horizontal or vertical.}
\medskip
 
Without loss of generality, let $v_3$ be in quadrant 4 (and $-v_3$ in quadrant 2) and let $v_4$ be in quadrant 3 (and $-v_4$ in quadrant 1).
 
\smallskip
\noindent\emph{Subcase 1: $\frac{m_1}{k}<a<m_1-\frac{m_1}{k}$ or $\frac{m_2}{k}<b<m_2-\frac{m_2}{k}$.}
\smallskip
 
Without loss of generality, let $\frac{m_1}{k}<a<m_1-\frac{m_1}{k}$ and $b>m_2-\frac{m_2}{k}$.
Then, repeatedly apply $v_3$ if $a<\frac{m_1}{2}$ and $v_4$ otherwise; since the $x$ components of $v_3$ and $v_4$ are bounded by $\frac{3m_1}{k}$,
these operations cannot push $a$ out of the range $\frac{m_1}{k}\ldots m_1-\frac{m_1}{k}$.
Since $b$ decreases with each operation, eventually $b$ will be within $\frac{m_2}{k}\ldots m_2-\frac{m_2}{k}$.
 
\smallskip
\noindent\emph{Subcase 2: $a$ isn't between $\frac{m_1}{k}\ldots m_1-\frac{m_1}{k}$ and $b$ is not between $\frac{m_2}{k}\ldots m_2-\frac{m_2}{k}$.}
\smallskip
 
Without loss of generality, let $a<\frac{m_1}{k}$ and $b>m_2-\frac{m_2}{k}$.
Then, repeatedly subtracting $v_4$ will either push $(a,b)$ into the middle region, in which case we are done, or into the zone where Subcase~1 can be used.
\end{proof}

\begin{lemma}
\label{lem:newperiod}
Given $v_1$ and $v_2$, if they lie in the same or opposite quadrants, then a period lying in an orthogonal quadrant can be constructed whose dimensions are bounded by $\frac{3m_1}{k}\times \frac{3m_2}{k}$.
\end{lemma}

\begin{proof}
Let $v_1$ and $v_2$ be the periods, in the same quadrant, without loss of generality quadrant 1
(if they lie in opposite quadrants, first negate one of them).
Potentially invert $v_2$ to ensure its $(1,1)$ direction component is opposite from $v_1$'s.
 
Repeatedly do the following with the current period $v_3$, starting from $0$, until $v_3$ lies in $Q_2$ or $Q_4$:
if $v_3$'s $(1,1)$ component matches $v_1$'s, add $v_2$ to it; otherwise, add $v_1$.
 
$v_3$'s $(1,1)$ component is bounded by the components of $v_1$ and $v_2$.
Since $v_1\neq\pm v_2$, over the process, $v_3$'s $(1,-1)$ component will eventually exceed such a bound, putting $v_3$ in $Q_2$ or $Q_4$.
 
\emph{Bound on $v_3$'s size:} before the final application of $v_1$ or $v_2$, $v_3$'s $(1,-1)$ component was bounded by its $(1,1)$ component,
which is bounded by $v_1$'s or $v_2$'s $(1,1)$ component. Then $v_1$ or $v_2$ is applied once. Hence $v_3$ is the sum of $3$ vectors bounded by
$\frac{m_1}{k}\times\frac{m_2}{k}$, so it is bounded by $\frac{3m_1}{k}\times\frac{3m_2}{k}$.
\end{proof}

If this new period $v_3$ is a period of the entire pattern, then by lemma 3.2, the pattern is lattice periodic with respect to $v_1$ and $v_2$. Otherwise, since the middle region of the pattern is periodic with respect to $v_3$ due to it being lattice periodic, and the whole pattern is not, the period breaker algorithm can be used.

\begin{lemma}
\label{lem:breakerresidue}
If the period-breaker algorithm terminates with multiple shifts remaining - which implies the existence of a horizontal or vertical period - then the pattern is lattice periodic with respect to that horizontal or vertical period together with some other period.
\end{lemma}

\begin{proof}
Since there is a horizontal period, and since the pattern is 2D periodic there must be some other distinct period, lemmas 3.1 and 3.2 can be applied to show the pattern is lattice periodic.
\end{proof}

\subsubsection{Deterministic Sampling for a Lattice Periodic Pattern}
 
Let the current period vectors be $v_1$ and $v_2$. At every iteration, we perform the following.
 
Find one surviving shift, $S_{\mathrm{init}}=(X_{\mathrm{init}},Y_{\mathrm{init}})$, and let $S_c=(X_c,Y_c)=\bigl(\frac{m_1}{2k},\frac{m_2}{2k}\bigr)$.
Add $S_c-S_{\mathrm{init}}$ to each column in the deterministic sample, taken modulo $v_1$ and $v_2$ so that they remain middle columns.
This translates the surviving shift to $(X_c,Y_c)$.
 
Search for a surviving shift not on the lattice defined by $(X_c,Y_c)$, $v_1$, and $v_2$.
If none exists, deterministic sampling is complete. If another exists, denote it $(X_o,Y_o)=S_o$.
 
If a column exists where $S_c$ and $S_o$ differ (if there exists one, there must be a middle column due to lattice periodicity),
add it to the deterministic sample, randomly choosing a character from $S_c$ or $S_o$.
Otherwise, $v_3=(X_c-X_o,\;Y_c-Y_o)$, which is not a linear combination of $v_1$ and $v_2$, is a period of the pattern.
 
Note that linear combinations of $v_1$ and $v_2$ tile the plane in parallelograms, and so $v_3$ must lie in or on the edge of some such parallelogram.
Let its corners be $v_{P_1}$, $v_{P_2}=v_{P_1}+v_1$, $v_{P_3}=v_{P_1}+v_2$, and $v_{P_4}=v_{P_1}+v_1+v_2$, where all corners are linear combinations of $v_1$ and $v_2$.
 
Divide the parallelogram into $4$ similar ones by dividing across the lines joining the midpoints of opposite sides.
$v_3$ will lie in or on the edge of one; without loss of generality, let it be the parallelogram with $v_{P_1}$ as a vertex.
Then, the triangles defined by the vectors $(v_3-v_{P_1})$ and $\frac{v_1}{2}$, as well as $(v_3-v_{P_1})$ and $\frac{v_2}{2}$, form disjoint regions within the parallelogram.
If the former triangle is smaller in size, make $(v_3-v_{P_1})$ and $v_1$ the new periods; otherwise, make them $(v_3-v_{P_1})$ and $v_2$.
 
\begin{lemma}
\label{lem:halving}
The parallelograms formed by the new period and the selected current period has area at most half that of the original parallelogram.
\end{lemma}
 
\begin{proof}
Without loss of generality, assume the new vectors chosen are $v_1$ and $v_3-v_{P_1}=v_\Delta$.
Since the triangle defined by $\frac{v_1}{2}$ and $v_\Delta$ is at most half of a quarter of the area of the parallelogram defined by $v_1$ and $v_2$,
the parallelogram defined by $\frac{v_1}{2}$ and $v_\Delta$ is at most $\frac14$ the area of the parallelogram defined by $v_1$ and $v_2$.
Hence the parallelogram defined by $v_1$ and $v_\Delta$ has area at most half that of the parallelogram defined by $v_1$ and $v_2$.
\end{proof}
 
We then adopt this current period together with the new period as the new pair of period vectors. By Lemma~\ref{lem:halving}, each iteration either the size of the parallelogram formed by $v_1$ and $v_2$ is halved or the total shifts (modulo $v_1$ and $v_2$) is halved, keeping the total iterations and the deterministic sample size $\widetilde{O}(1)$.

\subsubsection{Period-Breaker Algorithm}
Note: If this stage is reached, any previously constructed deterministic sample should be disregarded.

This algorithm handles the following situation. Suppose there 
is $(\delta_x, \delta_y)$ where $|\delta_x| \leq 3m_1/k$ and $|\delta_y| \leq 3m_2/k$, such that:

\begin{itemize}
\item the middle region of the pattern is periodic with respect to $(\delta_x,\delta_y)$. This can be implied from there being \emph{no} middle column $(a,b)$ at which some 2 shifts $(x_1,y_1)$ and $(x_2,y_2)$ differ, where $(x_2-x_1,y_2-y_1)$ = $(\delta_x,\delta_y)$; but
\item there \emph{is} some pattern index $(c,d)$ with $P[c][d] \neq P[c + \delta_x][d + \delta_y]$.
\end{itemize}

Define a pattern index $(x,y)$ to be:
\begin{itemize}
\item \emph{good} if $P[x][y] = P[x + \delta_x][y + \delta_y]$,
\item \emph{bad} if $P[x][y] \neq P[x + \delta_x][y + \delta_y]$,
\item \emph{neutral} if $P[x + \delta_x][y + \delta_y]$ is out of range.
\end{itemize}

\begin{lemma}
If there is an $\frac{m_1}{k} \times \frac{m_2}{k}$ rectangle in the pattern, $R$, containing all good cells except exactly one row or one column containing at least 1 bad cell, a deterministic sample can be created leaving a horizontal or vertical line of at least one shifts.
\end{lemma}
\begin{proof}
(proof). Let the top left corner of the rectangle $R$ be at index $(X_r,Y_r)$, and let one bad index be at $(X_b,Y_b)$.
Let $R$ and the index in $R$ shifted by $(\delta_x,\delta_y)$ be as in the period-breaker setting; in particular $R$ contains no neutral cells.
 
Consider the columns $\bigl(X_r+\frac{m_1}{k},\,Y_r+\frac{m_2}{k}\bigr)$ and $\bigl(X_r+\frac{m_1}{k}+\delta_x,\,Y_r+\frac{m_2}{k}+\delta_y\bigr)$.
For any shift $(a,b)$, these store the indices $P\bigl[X_r+\frac{m_1}{k}-a\bigr]\bigl[Y_r+\frac{m_2}{k}-b\bigr]$ and
$P\bigl[X_r+\frac{m_1}{k}-a+\delta_x\bigr]\bigl[Y_r+\frac{m_2}{k}-b+\delta_y\bigr]$.
Note that these are an index in $R$ and the index shifted by $(\delta_x,\delta_y)$, respectively.
 
Make the deterministic sample columns $\bigl(X_r+\frac{m_1}{k},\,Y_r+\frac{m_2}{k}\bigr)$ and $\bigl(X_r+\frac{m_1}{k}+\delta_x,\,Y_r+\frac{m_2}{k}+\delta_y\bigr)$
with characters $P[X_b][Y_b]$ and $P[X_b+\delta_x][Y_b+\delta_y]$.
Since these are the characters stored by the shift $\bigl(X_r+\frac{m_1}{k}-X_b,\;Y_r+\frac{m_2}{k}-Y_b\bigr)$, denoted $S_0=(X_0,Y_0)$,
at least one shift survives this deterministic sample. Additionally, these characters are distinct, as $(X_b,Y_b)$ is bad.
 
Without loss of generality, let bad indices only exist in the row of $R$ with $Y$ indices $Y_b$.
Then, for any shift with $y$-coordinate not equal to $Y_0=Y_r+\frac{m_2}{k}-Y_b$, the index the first column is mapped to
cannot have $y$ index $Y_b$, and therefore must be good. Hence it will store the same character in the $2$ columns and be deleted.
As such, at most the horizontal line of shifts with $y$ coordinate $Y_0$ will survive.
 
From here, deterministic sampling can proceed analogously to the $1$-D case.
\end{proof}

Any index in the inner middle region --- i.e.\ with $4m_1/k \le x \le m_1 - 4m_1/k$ and $4m_2/k \le y \le m_2 - 4m_2/k$ --- is automatically good, due to the middle region being periodic with respect to $(\delta_x,\delta_y)$. Since a bad index exists by hypothesis, at least one bad index lies outside the inner middle region. The remainder of the pattern (outside the inner middle region) partitions into four \emph{cardinal strips} immediately adjacent to the inner middle region (above, below, left, right of it) and four \emph{corner regions} (the four corners not in any cardinal direction from the inner middle region). We split into two cases based on where the bad indices lie.

\medskip
\noindent\textbf{Case 1: There exists a bad index inside one of the four cardinal strips.}
\medskip

Whether this condition holds can be tested with a Dürr–Høyer search over each strip.

Without loss of generality, suppose a bad index lies in the \emph{top} cardinal strip, i.e.\ satisfies $4m_1/k \le x \le m_1 - 4m_1/k$ together with $y \ge m_2 - 4m_2/k$ (so it lies to the right of the inner middle region's right edge, above the region). Use Dürr–Høyer search to find, among such indices, the bad index with smallest $y$-coordinate; call it $(x_b, y_b)$.

If $x_b \le m_1/2$, the rectangle with top left corner at $(x_b,y_b)$ satisfies the condition for the deterministic sample, as it only has bad indices in its top row; otherwise the rectangle with top right corner at $(x_b,y_b)$ can be used.

\medskip
\noindent\textbf{Case 2: No bad index lies in any cardinal strip (so every bad index lies in a corner region).}
\medskip

Without loss of generality, assume a bad index lies in the top-left corner region, i.e.\ satisfies $0 \le x \le 4m_1/k$ and $0 \le y \le 4m_2/k$. Use Dürr–Høyer search to find, among such bad indices, the one with minimum $y$-coordinate, tiebroken by maximum $x$-coordinate, $x_b,y_b$. This selection ensures that the region $P[x_b:x_b+\frac{m_1}{k}][y_b:y_b+\frac{m_2}{k}]$ contains only good indices other than $(x_b,y_b)$, and therefore can be used as the rectangle to create the deterministic sample.

\begin{lemma}[Complexity of Deterministic Sampling]
\label{lem:samplingcomplexity}
The deterministic sampling process runs in $\widetilde O(\sqrt{M})$ time.
\end{lemma}

\begin{proof}

Each form of deterministic sampling takes $\widetilde{O}(1)$ iterations, and each iteration utilizes a constant number of $\widetilde{O}(\sqrt{M})$ Grover's or Dürr–Høyer queries. The period breaker algorithm also uses a constant number of $\widetilde{O}(\sqrt{M})$ Grover's or Dürr–Høyer queries followed by $\widetilde{O}(1)$ iterations of a constant number of $\widetilde{O}(\sqrt{M})$ Grover's or Dürr–Høyer queries. Hence, the total time complexity of deterministic sampling is $\widetilde{O}(\sqrt{M})$.

\end{proof}

\section{Ricochet Property with the Deterministic Sample}

The ricochet property states that once a deterministic sample matches at a given text position, the set of other positions in a surrounding neighborhood that can also match is tightly constrained. Which constraint applies depends on the periodicity type of $P$. We treat the two cases separately.

\subsection{Single Shift or Line of Shifts Surviving}
Partition the initial surviving-shift region into four quadrants of (roughly) equal size. Without loss of generality, suppose the shift surviving deterministic sampling (or, in the 1-D periodic case, one of the arbitrary surviving shifts) lies in the bottom-right quadrant. Then in the $m_1/(2k) \times m_2/(2k)$ region to the bottom-right of a deterministic-sample match, the pattern cannot match --- except, in the 1-D periodic case, at translates of other surviving shifts by the period.

\subsection{Lattice Periodic Deterministic Sample}

Since a surviving shift sits at the center of the initial shift region, if the deterministic sample matches at a given position, then within a surrounding block of size $m_1/(2k) \times m_2/(2k)$, the pattern cannot match at any index that is not an integer linear combination of the periods away from the deterministic-sample match.

\section{The Skyline}

Consider a rectangle $R[0 \ldots r_1-1][0 \ldots r_2-1]$, where without loss of generality $r_1 < r_2$. Suppose we are given a binary function on $(i,j)$ with $0 \le i \le r_1$ and $0 \le j \le r_2$, assigning each index a value in $\{0, 1\}$. Call indices mapped to $1$ \emph{good} and indices mapped to $0$ \emph{bad}.

The \emph{skyline} aims to answer, for each row $l$ from $0$ to $r_1 - 1$, the maximum $k$ such that $R[0 \ldots l][0 \ldots k]$ contains only good indices. Formally, the skyline is the sequence $(l, k_l)$ such that $f(R[i][j]) = 1$ for all $i \le l$ and all $j \le k_l$.

We solve this recursively. At every stage of the recursion we construct the skyline on a sub-rectangle parameterized by $x_{\min}, x_{\max}, y_{\min}, y_{\max}$, with $x = x_{\max} - x_{\min} + 1$ rows and $y = y_{\max} - y_{\min} + 1$ columns. Initially $x_{\min} = 0,\, x_{\max} = r_1 - 1,\, y_{\min} = 0,\, y_{\max} = r_2 - 1$.

Let $xy$, the total number of cells relevant to the current portion of the skyline construction (within $x_l, x_r, y_b, y_t$), be the \emph{area}. We would like to decompose the skyline into two pieces of (roughly) equal area. We do this by binary searching for the $x$-coordinate of a ``midpoint'' --- initialized at $(x_l + x_r)/2$ --- that splits the skyline into two equal-area regions.

At each step of the binary search, for the trial $x$-coordinate $x_t$ we use Grover's algorithm (Dürr–Høyer) to find the bad index of least $y$-coordinate, $y_t$, within the region $R[x_{\min} \ldots x_t][y_{\min} \ldots y_{\max}]$. If no such index exists, we set $y_t := y_{\max} + 1$. The two resulting half-areas are $(x_t - x_{\min})(y_{\max} - y_t)$ and $(x_{\max} - x_t)(y_t - y_{\min})$; binary search equalizes them.

Once the correct $x$-coordinate of the midpoint is found, a second binary search --- this time on the $y$-coordinate --- refines the area balance exactly.

\begin{lemma}[Area Bound]
\label{lem:area}
If both halves have exactly equal area, each must have at most $\tfrac{1}{4}$ of the area of the larger enclosing region.
\end{lemma}

\begin{proof}
Let the left region have dimensions $a \times b$ and the right region have dimensions $c \times d$, where the enclosing region has dimensions $(a+c) \times (b+d)$ and $ab = cd$. It suffices to show
\begin{align*}
2(ab + cd) \le (a+c)(b+d) \;&\Longleftrightarrow\; ab + cd \le ad + bc \;\Longleftrightarrow\; (a - c)(d - b) \ge 0 \\
&\Longleftrightarrow\; \bigl(a - ab/d\bigr)(d - b) \ge 0 \;\Longleftrightarrow\; a\!\left(\tfrac{d-b}{d}\right)\!(d-b) \ge 0 \\
&\Longleftrightarrow\; \tfrac{a}{d}(d-b)^2 \ge 0,
\end{align*}
which is clearly true since $a$ and $d$ are positive and $(d-b)^2 \ge 0$.
\end{proof}

By the intermediate value theorem, some splitting point of this form must approximately exist. However, if one dimension of the region becomes too small, the split need not be close to half. To handle this edge case, if either dimension falls below some constant, we instead construct the skyline manually by locating the lowest violator in each row (or column) along the smaller dimension and tracking the jumps of the resulting skyline; this involves $O(1)$ Grover searches over a region of size the (larger) remaining dimension, so the base case also costs $\widetilde O(\sqrt{xy})$, consistent with the recurrence below.

\begin{lemma}[Skyline Construction Cost]
\label{lem:recurrence}
Constructing the skyline for a block of area $A = r_1 r_2$ costs $\widetilde O(\sqrt{A})$.
\end{lemma}

\begin{proof}
At a node of the recursion with region area $A$, the two binary searches (for $x_t$ then $y_t$) each take $O(\log A)$ steps, and each step is one Grover search over a region of size at most $A$, costing $O(\sqrt{A})$. So the cost at a single node, excluding its recursive calls, is $O(\sqrt{A} \log A)$. By Lemma~\ref{lem:area}, the node recurses into $2$ children each of area at most $A/4$, giving
\[
T(A) \;=\; O(\sqrt{A}\,\log A) \;+\; 2\, T(A/4).
\]
At recursion depth $t$ there are at most $2^t$ nodes, each of area at most $A/4^t$, so the total cost at level $t$ is
\[
2^t \cdot O\!\left(\sqrt{A/4^t}\,\log(A/4^t)\right) \;=\; O\!\left(\sqrt{A}\,\log(A/4^t)\right) \;=\; O(\sqrt{A}\,\log A).
\]
The recursion bottoms out after at most $t = O(\log A)$ levels, so summing across all levels,
\[
T(A) \;=\; O(\sqrt{A}\,\log^2 A) \;=\; \widetilde O(\sqrt{A}).
\]
\end{proof}

The final stage of the algorithm yields a series of registers encoding ``right'' or ``down'' moves of specified magnitudes. This representation is then converted into per-column heights (one per column, indexed by $r_1$), so that the total information stored is at most $\widetilde O(r_1) \le \widetilde O(\sqrt{r_1 r_2})$.

\section{Pattern Matching in the Text: Four-Corner Skyline Verification}
\label{sec:fourcorner}

Three of the subsequent pattern-matching subsections share a common verification step --- running four skylines based from the four corners of adjacent blocks, computing four cardinal-direction extreme bad indices, screening a shared inner region, and then Grover-searching for a candidate shift that satisfies all the constraints. We present that shared machinery once here and reference it from each case.

Fix a block $(i,j)$. Suppose we are given:

\begin{itemize}
\item a notion of ``good'' and ``bad'' text cells supplied by a $\widetilde O(1)$ oracle (whose definition differs per case --- see below), and
\item a residue condition that constrains which shifts are candidates (``in $\mathcal{T}$,'' ``in the surviving class,'' or ``in the lattice coset through $(x_m, y_m)$'' depending on the case).
\end{itemize}

\smallskip
\noindent\textbf{Four-corner skylines.}
Construct skylines of good cells based from, respectively, the bottom-right corner of block $(i,j)$, the bottom-left corner of block $(i+2k, j)$, the top-right corner of block $(i, j+2k)$, and the top-left corner of block $(i+2k, j+2k)$. These four skylines are independent of each other and of the four cardinal extremes defined below.

\smallskip
\noindent\textbf{Four cardinal-direction bad-index extremes.}
Independently from the four skylines, use Dürr–Høyer search over the four cardinal strips surrounding the current block to extract four bad-index extremes:
\begin{itemize}
\item $(x_u, y_u)$: the bad index of \emph{minimum $y$-coordinate} in the top strip $T[(i+1)m_1/2k \ldots i\,m_1/2k + m_1][\, j\,m_2/2k \ldots (j+1)m_2/2k\,]$;
\item $(x_D, y_D)$: the bad index of \emph{maximum $y$-coordinate} in the bottom strip $T[(i+1)m_1/2k \ldots i\,m_1/2k + m_1][\, j\,m_2/2k + m_2 \ldots (j+1)m_2/2k + m_2\,]$;
\item $(x_l, y_l)$: the bad index of \emph{maximum $x$-coordinate} in the left strip $T[\, i\,m_1/2k \ldots (i+1)m_1/2k\,][(j+1)m_2/2k \ldots j\,m_2/2k + m_2]$;
\item $(x_r, y_r)$: the bad index of \emph{minimum $x$-coordinate} in the right strip $T[\, i\,m_1/2k + m_1 \ldots (i+1)m_1/2k + m_1\,][(j+1)m_2/2k \ldots j\,m_2/2k + m_2]$.
\end{itemize}
(Here we have renamed the bottom-strip extreme to $(x_D, y_D)$ to avoid clashing with the period-breaker gap $(\delta_x, \delta_y)$.)

\noindent\textbf{Shared inner-region bad-cell screening.}
Use Grover's search to test whether the shared region
\[
T[\,(i+1)m_1/2k \ldots i\,m_1/2k + m_1\,][\,(j+1)m_2/2k \ldots j\,m_2/2k + m_2\,]
\]
contains any bad cell. Since this region is contained in every pattern instance whose top-left corner lies in the current block, the presence of a bad cell here rules out every candidate in the block; otherwise this region can be disregarded entirely.

\smallskip
\noindent\textbf{Per-candidate verification oracle.}
For each pattern instance with top-left corner at $(x, y)$ satisfying the residue condition of the case, the full pattern instance consists only of good cells if and only if:
\begin{itemize}
\item $(x, y)$ lies within the top-left-corner skyline,
\item $(x + m_1, y)$ lies within the top-right-corner skyline,
\item $(x, y + m_2)$ lies within the bottom-left-corner skyline,
\item $(x + m_1, y + m_2)$ lies within the bottom-right-corner skyline,
\item $x_l < x < x_r$, and
\item $y_u < y < y_D$.
\end{itemize}
This conjunction can be checked by an $O(1)$ oracle.

\smallskip
\noindent\textbf{Final Grover search.}
Use Grover's search over candidate shifts, filtered by the residue condition of the case, returning any candidate for which the above conjunction holds. A pattern match in the block exists if and only if some candidate is returned.

\begin{figure}[t]
    \centering
    \includegraphics[width=0.65\textwidth]{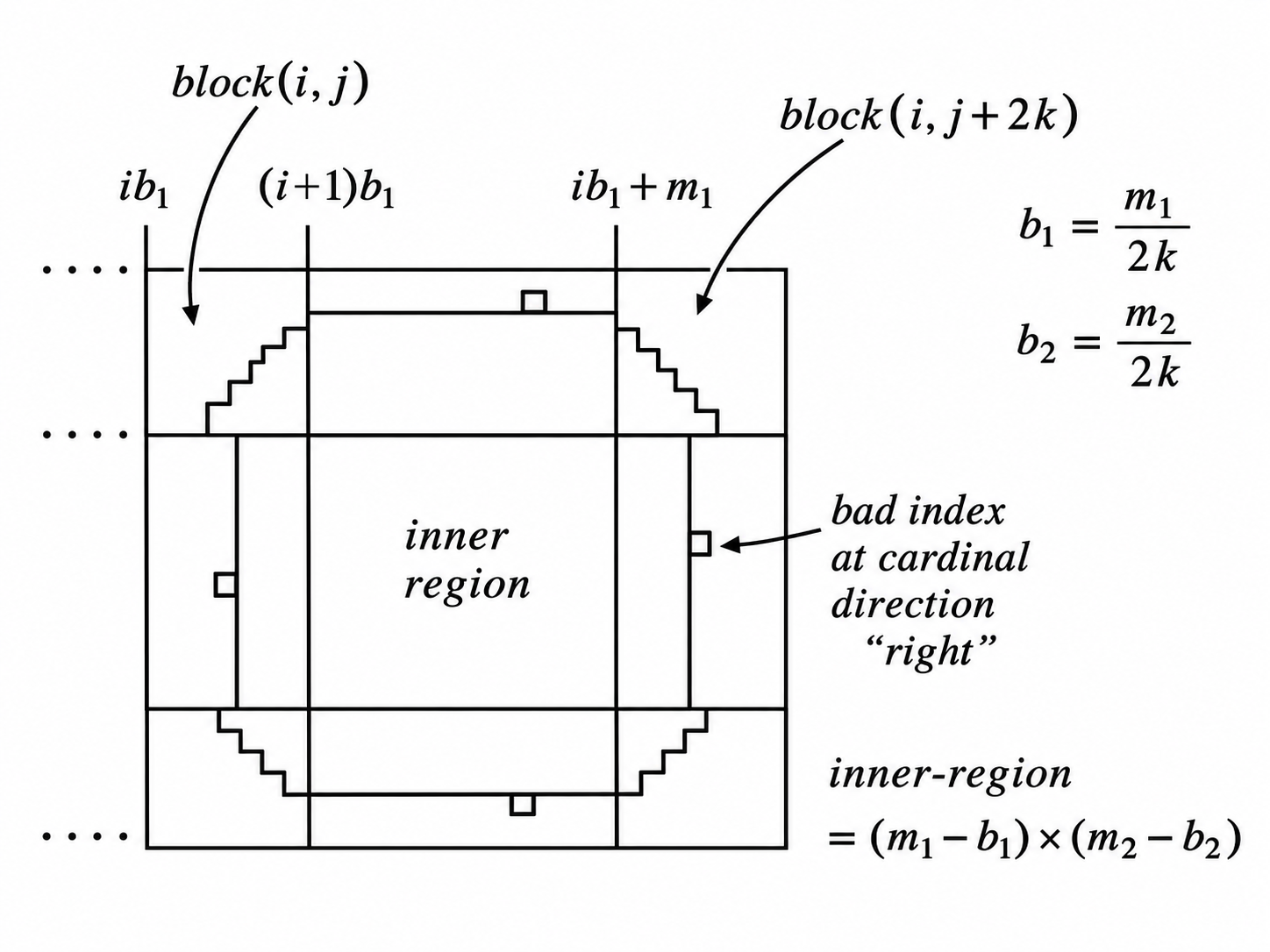}
    \caption{4 Skylines and extended regions for block $(i,j)$}
    \label{fig:block-partition}
\end{figure}

\smallskip

\subsection{Pattern Matching if no Other Shifts Remain in the Ricochet Zone of a Surviving Shift}

Divide the text into blocks of size $n_1 / (2 k m_1) \times n_2 / (2 k m_2)$.

For each block $(i, j)$, use the skyline algorithm to find a skyline of the region $T[i\,m_1 \ldots (i+1)m_1][\,j\,m_2 \ldots (j+1)m_2\,]$ relative to the top-left corner that contains no deterministic-sample matches. Then, by the ricochet property, the only potential pattern-match positions are among the at most $O(\sqrt{M})$ deterministic-sample matches lying just outside corners of the skyline.

We then maintain a list $\mathcal{T}$ of candidate match positions, sorted by increasing $x$-coordinate, initially empty. For each deterministic-sample-match position $(x, y)$ encountered along the skyline, proceed as follows:
\begin{enumerate}
\item If $\mathcal{T}$ is empty, add $(x, y)$ to it and continue to the next match along the skyline.
\item Otherwise, let the last element of $\mathcal{T}$ be $(x_p, y_p)$. Use the witness table to either find an index $(x_d, y_d)$ in the text at which pattern instances based at $(x, y)$ and at $(x_p, y_p)$ would disagree, or determine that no such index exists.
\item If no such $(x_d, y_d)$ exists (either in $\mathcal{T}$ or in the witness table), add $(x, y)$ to $\mathcal{T}$ and continue to the next match along the skyline.
\item If the pattern instance at $(x, y)$ matches the text at $(x_d, y_d)$, remove $(x_p, y_p)$ from $\mathcal{T}$ and restart from step~1. Otherwise, continue to the next match along the skyline.
\end{enumerate}

By construction, only positions in $\mathcal{T}$ can match the pattern, and no two pattern instances based at positions in $\mathcal{T}$ disagree at any text index.

Build, in $\widetilde O(\sqrt{M})$ time, a query structure that given $x$- and $y$-coordinate ranges returns an index into $\mathcal{T}$ falling in that range; queries cost $\widetilde O(1)$.

Define a $\widetilde O(1)$ oracle labeling each text index as \emph{good} or \emph{bad}: good indices match the pattern instance being considered; bad indices either do not match or are irrelevant to any instance under consideration.

The oracle acts on a text index $(x_i, y_i)$ as follows. First, query $\mathcal{T}$ for an element $(x_r, y_r)$ with $x_i - m_1 \le x_r \le x_i$ and $y_i - m_2 \le y_r \le y_i$; this ensures the pattern instance based at $(x_r, y_r)$ covers $(x_i, y_i)$. If no such $(x_r, y_r)$ exists, label $(x_i, y_i)$ bad. Otherwise, label $(x_i, y_i)$ good if $T[x_i][y_i] = P[x_i - x_r][y_i - y_r]$ and bad otherwise.

Apply the four-corner skyline verification from Section~\ref{sec:fourcorner} with this good/bad oracle and the residue condition ``$(x, y) \in \mathcal{T}$'' to detect any pattern match in the block.

\subsection{Pattern Matching if the Pattern is 1-D Periodic and Another Shift Survives in the Ricochet Zone}

Let the pattern's period be $v = (x_v, y_v)$. Without loss of generality, assume $m_1 \le \sqrt{M} \le m_2$.

For each block $(i,j)$, use the skyline algorithm to find a skyline of $T[i\,m_1 \ldots (i+1)m_1][\,j\,m_2 \ldots (j+1)m_2\,]$ relative to the top-left corner that contains no deterministic-sample matches.

Call two text indices members of the same \emph{class} if they differ by an integer multiple of $v$. Since the pattern is only 1-D periodic, no two indices from distinct classes could support pattern instances whose overlapping region is free of conflicts.

Two cases arise, depending on $x_v$. In each we identify a single class that could contain all possible pattern matches.

\subsubsection{Finding the Surviving Class if $x_v = 0$}

Each class that could contain a pattern match must include a deterministic-sample match on the skyline together with the shifts lying at integer multiples of $y_v$ below it. There are at most $O(\sqrt{M})$ such classes. All indices within one such class share an $x$-coordinate.

We reduce these classes down to one surviving class via repeated pairwise \emph{duels}. In each duel, between class~1 (with $x$-coordinate $x_1$) and class~2 (with $x$-coordinate $x_2$), assume WLOG $x_1 < x_2$. Let $i_1$ be the index in class~1 of lowest $y$-coordinate, and $i_2$ the index in class~2 of highest $y$-coordinate. Since --- by the skyline's structure --- the $y$-coordinate of $i_2$ exceeds the $y$-coordinate of every index in class~1, the overlap region of the pattern instances at $i_1$ and $i_2$ sits entirely within every pattern instance based at an index in class~1 or class~2, with one potential exception: $i_3$, the index of class~2 with lowest $y$-coordinate, if $y(i_3) < y(i_1)$.

Using the witness table, find an index at which the pattern instances at $i_1$ and $i_2$ disagree. If $i_2$ matches the text at this disagreement index, then no index in class~1 can match the pattern, and class~2 wins the duel. Otherwise, $i_1$ matches at the disagreement; the only candidate in class~2 that might still match is $i_3$. Then duel $i_3$ against the index in class~1 of highest $y$-coordinate: since the overlap region of these two pattern instances lies inside every pattern instance in class~1, if $i_3$ matches at the disagreement index, class~2 wins; otherwise class~1 wins.

\subsubsection{Finding the Surviving Class if $x_v \neq 0$}

For every $x$-coordinate $x_i$ present in the block, consider the deterministic-sample match $(x_m, y_m)$ on the skyline with the greatest $x$-coordinate not exceeding $x_i$ (if no such match exists, every index in the block with $x$-coordinate $x_i$ lies outside the skyline and cannot match the pattern). Every index in the block with $x$-coordinate $x_i$ is either in the ricochet zone of $(x_m, y_m)$, or outside the skyline and hence not a pattern-match candidate.

Hence, the only index with $x$-coordinate $x_i$ that could match the pattern is the at-most-one index that is an integer multiple of $v$ away from $(x_m, y_m)$, i.e.\ $\bigl(x_i,\; y_m + \tfrac{y_v}{x_v}(x_i - x_m)\bigr)$, and only when $y_m + \tfrac{y_v}{x_v}(x_i - x_m)$ is an integer.

Since there is only one possible surviving index per $x$-coordinate, at most $O(\sqrt{M})$ surviving indices remain overall.

Maintain a list $\mathcal{T}$ of candidate indices, sorted by increasing $x$-coordinate, initially empty. At all times $\mathcal{T}$ contains only indices whose pattern instances pairwise have no index of disagreement, so $\mathcal{T}$ can contain indices from at most one class.

For each surviving index $(x_i, y_i)$, in order of increasing $x_i$, repeatedly do:
\begin{enumerate}
\item If $\mathcal{T}$ is empty or already contains elements from $(x_i, y_i)$'s class, add $(x_i, y_i)$ to $\mathcal{T}$.
\item Otherwise, duel $(x_i, y_i)$ against the last element of $\mathcal{T}$ by consulting the witness table's disagreement index. If $(x_i, y_i)$ matches the text at the disagreement index, remove the last element of $\mathcal{T}$ and repeat; otherwise skip $(x_i, y_i)$.
\end{enumerate}

In the end, $\mathcal{T}$ contains indices from at most one class --- the surviving class.

\subsubsection{Pattern Matching Once the Surviving Class Is Known}

Let $(x_r, y_r)$ be any index in the surviving class.

Define a $\widetilde O(1)$ oracle labeling text cells as \emph{good} or \emph{bad}: good indices match the pattern instances under consideration; bad indices either do not match or are irrelevant.

To implement the oracle for query index $(x, y)$: find an integer $a$ (and in the $x_v \ne 0$ sub-case, this takes the form of using the period relation once) such that $(x, y) + a v = (x', y')$ lies inside $T[x_r \ldots x_r + m_1][y_r \ldots y_r + m_2]$, by using the intersection of the four inequalities $x_r < x',\; x' < x_r + m_1,\; y_r < y',\; y' < y_r + m_2$. If no such $a$ exists, $(x, y)$ is bad. Otherwise, $(x, y)$ is good iff $P[x' - x_r][y' - y_r] = T[x'][y']$.

Apply the four-corner skyline verification from Section~\ref{sec:fourcorner} with this oracle and with residue condition ``$(x, y)$ belongs to the same class as $(x_r, y_r)$'' to detect any pattern match in the block.

\subsection{Pattern Matching for a Lattice-Periodic Pattern}

Let the two periods be $v_1$ and $v_2$.

Use Grover's search to find some deterministic-sample match $(x_m, y_m)$. By the ricochet property, the pattern either matches at an integer linear combination of $v_1$ and $v_2$ away from $(x_m, y_m)$, or matches nowhere in the block.

Define a $\widetilde O(1)$ oracle labeling text cells as \emph{good} or \emph{bad}: good indices match the pattern instances under consideration; bad indices either do not match or are irrelevant.

To implement the oracle for query index $(x, y)$: find integers $a, b$ such that $(x, y) + a v_1 + b v_2 = (x', y')$ lies inside $T[x_m \ldots x_m + m_1][y_m \ldots y_m + m_2]$, using the intersection of the four inequalities $x_m < x',\; x' < x_m + m_1,\; y_m < y',\; y' < y_m + m_2$. Then $(x, y)$ is good iff $P[x' - x_m][y' - y_m] = T[x'][y']$.

Apply the four-corner skyline verification from Section~\ref{sec:fourcorner} with this oracle and with residue condition ``$(x, y)$ is an integer linear combination of $v_1$ and $v_2$ away from $(x_m, y_m)$'' to detect any pattern match in the block.

\section{Time Complexity}
\label{sec:complexity}

As shown earlier, the time complexity of deterministic sampling is $\widetilde{O}(\sqrt{M})$.

Per block, the cost for creating the 4 block skylines is shown to be $\widetilde{O}(\sqrt{M})$, and finding the nearest bad cells on the edges and checking if the shared region only contains good cells are all $\widetilde{O}(\sqrt{M})$ Grover's or Dürr–Høyer algorithm queries.

In the 1st case of pattern matching, per block, there is an $\widetilde{O}(\sqrt{M})$ cost of finding the initial skyline shifts, another $\widetilde{O}(\sqrt{M})$ to duel them to create $\cal{T}$, and a final $\widetilde{O}(\sqrt{M})$ cost to create a query structure for elements of $\cal{T}$ amounting to a $\widetilde{O}(\sqrt{M})$ additional cost.

In the 2nd case of pattern matching, per block, there is an additional $\widetilde{O}(\sqrt{M})$ cost to finding the skyline shifts and a $\widetilde{O}(\sqrt{M})$ cost to find the surviving class.

In the 3rd case of pattern matching, there is an additional $\widetilde{O}(\sqrt{M})$ cost per block to find a deterministic sample match.

Hence, the total cost per block in any case is $\widetilde{O}(\sqrt{M})$. Querying over the $O(N/M)$ blocks with Grover's search, the total complexity arrives to $\widetilde{O}(\sqrt{N/M}\times \sqrt{M}) = \widetilde{O}(\sqrt{N})$.

\begin{theorem}
Given a pattern $P[0..m_1-1][0..m_2-1]$ and a text $T[0..n_1-1][0..n_2-1]$ and a witness table as oracles, (where $m_1m_2 = M \le n_1n_2 = N$) finding whether $P$ occurs in $T$ can be found by $\widetilde{O}(\sqrt{N})$ quantum query complexity.
\end{theorem}
\qed

\noindent\textbf{AI disclosure:} Generative AI has been used for proof reading, fixing language issues, checking logical issues, creating pseudocodes and figures and creating latex from drafts. All the ideas and solutions are authors' own. AI generated parts were carefully reviewed for accuracy. Authors are fully responsible for the content.

\bibliographystyle{plain}

\end{document}